\documentclass{article}

\evensidemargin=\oddsidemargin
\def\Title#1#2#3{%
    \baselineskip=18pt
    \begin{center}
          {\large\bf{#1} \\ }
          \bigskip\bigskip
          {#2} \\
          {#3} \\
    \end{center}}
\long\def\Abstract#1{%
         \bigskip
         \parbox{0.93\textwidth}{%
                 \begin{center}
                       {\bf Abstract} \\
                 \end{center}
                 \medskip{\baselineskip=14pt #1}
                 \vss}
         \bigskip}

\makeatletter
\renewcommand{\section}%
 {\@startsection{section}{1}{0pt}%
  {-3.25ex plus -1ex minus -.2ex}{1.5ex plus .2ex}%
  {\vspace*{5mm}\raggedright\large\bf }}
\renewcommand{\subsection}%
 {\@startsection{subsection}{2}{0pt}%
  {-2.25ex plus -.5ex minus -.2ex}{-1.5ex plus -.2ex}{\bf }}
\renewcommand{\subsubsection}%
 {\@startsection{subsubsection}{3}{0pt}%
  {-1.25ex plus -.2ex minus -.1ex}{-1.2ex plus -.2ex}{\bf }}

\makeatother

\begin{document}

\Title{Wave function of the Universe, path integrals and gauge invariance}%
{T. P. Shestakova}%
{Department of Theoretical and Computational Physics,
Southern Federal University,\\
Sorge St. 5, Rostov-on-Don 344090, Russia \\
E-mail: {\tt shestakova@sfedu.ru}}

\Abstract{The paper is devoted to some of the difficulties which the Wheeler -- DeWitt quantum geometrodynamics encountered, in particular, a strong mathematical proof that this theory is gauge-invariant, the definition of the wave function of the Universe through a path integral and the illegality of asymptotic boundary conditions in quantum gravity, the derivation of the Wheeler -- DeWitt equation from the path integral and the equivalence of the Dirac quantization scheme with other approaches, the problem of definition of physical states in quantum gravity, possible realizations of the Everett concept of ``relative states''. The problems are rarely discussed in the literature. They are related with the guiding idea that quantum theory of gravity must gauge invariant. It will lead to the question if it is possible to achieve this goal in a mathematically consistent way.}

\section{Introduction}
The Wheeler -- DeWitt quantum geometrodynamics \cite{DeWitt} is the first and, probably, most known attempt to quantize gravity. But it is very different from all other field theories not only because until now we have not got enough observational data to verify it. Quantum geometrodynamics, where the main object is the wave function of the Universe, is, in a sense, a return to the ideology of quantum mechanics with its problem of interpretation. There exists an opinion that the notion of the wave function cannot be applied to the Universe as a whole.

It is clear that the analogy with quantum mechanics has influenced the essence of the Wheeler -- DeWitt geometrodynamics. While in quantum mechanics the wave function is a probability amplitude to find a particle in a point with given coordinates, in quantum geometrodynamics the wave function is a probability amplitude for the Universe to have some geometry with a given metric. It follows from the primary gravitational constraints that the wave function depends only on space metric $\gamma_{ij}$, and the momentum constraints imply that the wave function is invariant under three-dimensional coordinate transformations (it was firstly pointed out by Higgs \cite{Higgs}, the proof which is valid under the assumption that infinitesimal parameters vanish at infinity can be found in \cite{Kiefer1}). So, the interpretation of the wave function as the probability amplitude for a certain space geometry seems to be obvious.

Here we face the first question: How can this theory help us to understand the Very Early Universe? It was declared \cite{Hartle} that the wave function describes an initial state for the subsequent classical evolution of the Universe. But it contradicts to the fact that a solution to the Wheeler -- DeWitt equation does not depend on time and, therefore, the probability distribution determined by the wave function must be true at every moment of the Universe existence. Thus, if the probability distribution predicted that geometries with small values of the scale factor are more feasible, the Universe would have had no chance to become macroscopic. Otherwise, if the probability distribution predicted that large values of the scale factor are more probable, it would mean just the statement of the fact that our Universe is macroscopic, but it would not tell us anything about the early stage of its existence. If so, the Wheeler -- DeWitt theory can hardly add something to our understanding of the beginning of the Universe.

But the analogy with quantum mechanics goes even further. In fact, some authors state that spacetime does not exist in quantum gravity. For instance, Kiefer \cite{Kiefer2} wrote:
\begin{quote}
``In classical canonical gravity, a spacetime can be represented as a `trajectory' in configuration space -- the space of all three-metrics\ldots\ Since no trajectories exist anymore in quantum theory, no spacetime exists at the most fundamental, and therefore also no time coordinates to parameterize any trajectory.''
\end{quote}
The same idea one can find in the book by Rovelli \cite{Rovelli1}:
\begin{quote}
``\ldots in quantum gravity the notion of spacetime disappears in the same manner in which the notion of trajectory disappears in the quantum theory of a particle.''
\end{quote}
However, the structure of spacetime plays a significant role in general relativity. Do we really need to reject it when quantizing gravity?

Rovelli \cite{Rovelli2} called the Wheeler -- DeWitt equation ``a strange equation, full of nasty features'', but, in spite of all this, it is ``a milestone in the development of general relativity''. What gives us confidence that this equation is fundamental? How can we be so sure that it expresses gauge invariance of quantum gravity? The Wheeler -- DeWitt equation is the result of application of the Dirac quantization scheme \cite{Dirac1,Dirac2} to gravitational theory. Dirac just postulated that after quantization any constraint must become a condition on a state vector. But this postulate has been never confirmed by direct experiments, since very successful and experimentally verified gauge theories like quantum electrodynamics were based on other theoretical methods than the Dirac scheme.

Apparently, some physicists felt the lack of ample grounds for the Wheeler -- DeWitt quantum geometrodynamics and tried to reveal a firm foundation for its main equation making use of other approaches, in particular, those based on path integral quantization. In this paper I shall review some attempts to strengthen results of quantum geometrodynamics. In Section 2 the Wheeler -- DeWitt equation will be considered in the framework of canonical quantization, in Section 3 a definition of the wave function of the Universe through a path integral will be discussed. I shall describe derivation of the Wheeler -- DeWitt equation from different forms of the path integral in Section 4. Equivalence of various approaches will be analysed in Section 5. In section 6 I shall remind of attempts of a mathematical realization of the Everett ``relative states'' concept and present the main conclusions.

\section{The Wheeler -- DeWitt equation and canonical quantization}
DeWitt \cite{DeWitt} presented his famous equation in the form
\begin{equation}
\label{WDW}
\left(G_{ijkl}\frac{\delta}{\delta\gamma_{ij}}\frac{\delta}{\delta\gamma_{kl}}+\sqrt{\gamma}R^{(3)}\right)\Psi=0,
\end{equation}
\begin{equation}
\label{DWSM}
G_{ijkl}=\frac1{2\sqrt\gamma}\left(\gamma_{ik}\gamma_{jl}+\gamma_{il}\gamma_{jk}-\gamma_{ij}\gamma_{kl}\right),
\end{equation}
$\gamma$ is the determinant of the space metric $\gamma_{ij}$, $R^{(3)}$ is 3-curvature. $G_{ijkl}$ is often referred to as ``the DeWitt supermetric'', a contravariant metric on space of all three-dimensional metrics $\gamma_{ij}$. However, discussing operator ordering in the Wheeler -- DeWitt equation Hawking and Page pointed out that the supermetric depends on the lapse function $N$ \cite{HP}, namely, the covariant supermetric is $\displaystyle\frac1N G^{ijkl}$, where $G^{ijkl}$ is the inverse of (\ref{DWSM}). It implies that the form of the Wheeler -- DeWitt equation depends on a relation between $N$ and $\gamma_{ij}$. One would obtain a family of the Wheeler -- DeWitt equations corresponding to different relations; as Hawking and Page wrote, ``there is one Wheeler-DeWitt equation for each field $N$''. To avoid this ambiguity they proposed to regard $N$ as a field independent on $\gamma_{ij}$. It means that $N$ is a constant, say, for definiteness, $N=1$, and returns us to Eq. (\ref{WDW}).

But it also means imposing an additional condition on the gauge variable $N$. This only condition, of course, does not fix a reference frame completely, but it restricts its choice. In the case of minisuperspace cosmological models when $N$ is the only gauge degree of freedom, the reference frame is totally fixed by this condition. However, the point noticed by Hawking and Page has not got due attention. Yet DeWitt wrote that, if desired, one can always assign definite values to the lapse and shift functions and he used the choice $N=1$, $N_i=0$. So, DeWitt was sure in gauge invariance of the constructed theory. Nevertheless, the dependence of the form of the main equation on the choice of the additional condition for $N$ is a serious indication that the theory may not be invariant, and from rigorous mathematical point of view one should examine it.

An attempt to prove gauge invariance has been made by Barvinsky. In \cite{Barv} a covariant operator realization of the gravitational constraints was explored using geometrical structure of the superspace manifold. A metric on the superspace was defined by contracting the DeWitt supermetric with some lapse function. However, the lapse function cannot be chosen arbitrary. As was shown in \cite{Barv}, only under the choice $N=1$ a desired operator realisation can be obtained. It is the same choice for $N$ that was used by DeWitt and proposed by Hawking and Page. Note that gauge invariance implies that the whole theory including a procedure of derivation of the equations for the wave function of the Universe, which are gravitational constraints in an operator form, and, as a consequence, the wave function must not depend on a choice of the lapse and shift functions, $N$ and $N^i$. Thus, though the resulting equations seem to be covariant, the whole procedure of their derivation is not. In other words, if one made another choice for $N$, one would get another form of operator gravitational constraints.

I shall now turn to the path integral approach.

\section{Path integral definition of the wave function of the Universe}
The seminal paper by Hartle and Hawking \cite{HH} aroused interest in quantum gravity and cosmology. Based on the analogy with Feynman formulation of quantum mechanics they gave a definition of the wave function through a path integral. The starting point was the probability amplitude
\begin{equation}
\label{H-amp}
\langle g_2, \phi_2, S_2 |\, g_1, \phi_1, S_1\rangle
\end{equation}
to go from a state with a spacetime metric $g_1$ and matter fields $\phi_1$ on a hypersurface $S_1$ to a state with a spacetime metric $g_2$ and matter fields $\phi_2$ on a hypersurface $S_2$, as a sum over all field configurations $g$ and $\phi$. At this point one can notice the difference between the path integral approach and canonical quantization. In the latter only space metric is of importance while gauge degrees of freedom (the lapse and shift functions, or $g_{0\mu}$-components of metric tensor) are believed to be irrelevant, and in the former all degrees of freedom contribute to the path integral. Hawking criticized the canonical approach since it implies the split into one time and three spatial dimensions that seems to be contrary to the spirit of relativity, and one would expect that in quantum gravity all possible topologies of spacetime should be allowed \cite{Hawk}. So, spacetime maintains its significance in the path integral approach.

But in the Wheeler -- DeWitt theory the wave function depends only on space component of the metric. In other words, we would rather deal with the amplitude
\begin{equation}
\label{DW-amp}
\langle \gamma_2, \phi_2, S_2 |\, \gamma_1, \phi_1, S_1\rangle,
\end{equation}
where $\gamma_1$ denotes a space metric on the hypersurface $S_1$, etc. To reconcile the both approaches Hartle and Hawking discussed what was meant by fixing the 4-geometry on a given spacelike surface. The answer was that one should impose gauge conditions near the surface so that all gravitation degrees of freedom but 3-metric would be fixed. It is not clear, however, how to impose the gauge conditions in the path integral, since the path integral is taken of exponent of a classical action of gravitational and matter fields, and the classical action does not include a gauge-fixing term, in contrast with an effective action in the Faddeev -- Popov or Batalin -- Fradkin -- Vilkovisky (BFV) methods of quantization. It is also not obvious, if the amplitude or the wave function defined in this way is independent on the choice of the gauge conditions, although it was tacitly supposed. But is not it possible to interpret it as the amplitude to go to a state a space metric $\gamma_2$ and matter fields $\phi_2$ on $S_2$ {\it under} the imposed gauge condition?

One could also require that the following relation must hold true \cite{Hawk}:
\begin{equation}
\label{H-rel}
\langle \gamma_2, \phi_2, S_2 |\, \gamma_1, \phi_1, S_1\rangle
 =\sum \langle \gamma_2, \phi_2, S_2 |\, \gamma_3, \phi_3, S_3\rangle
 \langle \gamma_3, \phi_3, S_3 |\, \gamma_1, \phi_1, S_1\rangle,
\end{equation}
where summing is over all states on the intermediate surface $S_3$. It is reasonable to assume that summing over all states means integrating over all 4-metrics on $S_3$. Then no gauge conditions must be fixed on this hypersurface.

These subtle questions can be naturally resolved in more elaborated approaches to quantization of gauge theories. However, as we shall see in the next Section, other problems appear in these approaches.

The Wheeler -- DeWitt equation is obtained in \cite{HH} by demanding that the wave function must not depend on the lapse function $N$:
\begin{equation}
\label{H-WDW}
0=\int Dg D\phi \left[\frac{\delta S}{\delta N}\right] \exp\left(i S[g,\phi]\right).
\end{equation}
In fact, it gives the classical Hamiltonian constraint ${\cal H}=\displaystyle\frac{\delta S}{\delta N}=0$ that must be translated into an operator equation after making some choice of operator ordering. At this stage the path integral approach cannot offer a more rigorous derivation of the Wheeler -- DeWitt equation then the canonical formalism.

There are two important points in this consideration. Firstly, gauge degrees of freedom do contribute into the path integral. And, secondly, one needs to fix some gauge conditions that poses the question if the path integral is independent on their choice.

\section{Derivation of the Wheeler -- DeWitt equation from a path integral}
It is well-known that a path integral over gauge fields is divergent. In the approach by Hawking \cite{Hawk} the problem is supposed to be overcome by means of the Wick rotation (the rotation of the time axis in the complex plane and the replacement of the action by the so-called ``Euclidean'' action). However, in other approaches to quantization of gauge fields \cite{FP,Fadd,BFV1,BFV2,BFV3,BV} the action is replaced by an effective one which include a gauge fixing and ghost terms. All these approaches were proposed for non-gravitational theories and aimed at constructing an $S$-matrix expressed by a path integral.

The construction of the $S$-matrix, of course, answers to the main goal of ordinary quantum field theory, which is the calculation of probabilities of various processes in particle physics. The $S$-matrix gives a transition amplitude between asymptotic (``in'' and ``out'') states. It is assumed that the states are physical, i.e. they are gauge-independent and do not contain any ghost fields. Accordingly, the path integral for the $S$-matrix is considered under the so-called asymptotic boundary conditions: zero boundary conditions for ghosts and Lagrange multipliers of gauges.

In 1980s, when the interest in quantum geometrodynamics increased, some authors had a purpose to demonstrate equivalence of the Wheeler -- DeWitt theory to generally accepted methods of quantisation such as the Faddeev -- Popov or BFV approaches. The demonstration of equivalence includes, in particular, the procedure of derivation of the Wheeler -- DeWitt equation from a path integral with an appropriate effective action. As a matter of fact, the question if these approaches are applicable without any modification to quantization of the Universe as a whole has not been investigated.

Hartle and Hawking defined the wave function through a transition amplitude between two spacelike hypersurfaces. Meanwhile, states of gravitational and other fields on these hypersurfaces are not analogous to the asymptotic states of ordinary field theory. Particles in the ``in'' and ``out'' states are beyond an interaction region, but the gravitational field does not vanish, in general, on the mentioned hypersurfaces. It is especially clear for a closed universe, the model which is so preferred by cosmologists. The $S$-matrix problem can be posed in the gravitational theory only in the case of asymptotically flat spacetime, and even in this case one can speak about propagation of gravitation waves and matter fields on a fixed background but not about a complete theory of quantum gravity.

The role of ghost fields in the gravitational theory has not been understood yet, however, ghosts appear inside the interaction region and are required to save unitarity of the transition amplitude. Thus, one cannot conclude that the state on an arbitrary hypersurface is free from ghosts. Also, as we saw in the previous Section, one may need to impose some gauge conditions to determine a 4-geometry on the hypersurface. It means that the asymptotic boundary conditions removing ghosts and gauge fixing on the hypersurface are not justified in gravity. The question, what a physical state in quantum gravity is, still has no satisfactory answer. I shall return to this question in the next section.

Nevertheless, the asymptotic boundary conditions are used without doubts in many works on quantum geometrodynamics, for example, in \cite{BP1,BP2}. In \cite{BP2} the authors stated that quantum geometrodynamics is not a closed physical theory and its postulates should be deduced from more basic principles of quantum theory of gauge fields. The basic principles are those of path integral quantization of true physical degrees of freedom \cite{Fadd}, when the constraints
\begin{equation}
\label{constr}
T_{\mu}(q, p)=0
\end{equation}
are complemented by the so-called unitary gauges depending only on phase variables,
\begin{equation}
\label{u-gauge}
\chi^{\mu}(q, p)=0.
\end{equation}
Moreover, because of the coordinate representation of the wave function, it was proposed to consider gauge conditions depending only on coordinates $q$: $\chi^{\mu}(q)=0$.

Quantization in unitary gauges is well established for electromagnetic field when one uses the only constraint and one gauge condition to reduce four components of the vector potential to two physically meaningful degrees of freedom. The application of this approach is problematic for the isotropic cosmological model in which the scale factor is the only physical degree of freedom. The model can be complemented by a scalar field and the application of this method leads to the exclusion of all gravitational degrees of freedom from a physical description of the model. It shows that well-known and established approaches being applied to gravity can give results which are not typical for ordinary gauge theories and necessary to be understood.

Returning to the derivation of the Wheeler -- DeWitt equation, it has been demonstrated in \cite{BP1,BP2} that a solution to quantum gravitational constraints (and the Wheeler -- DeWitt equation among them) can be presented in the form
\begin{equation}
\label{BWF}
\Psi(q)=\int dq_0 Z(q, q_0)\varphi(q_0),
\end{equation}
where $Z(q, q_0)$ is the transition amplitude in the form of the Faddeev -- Popov path integral over the phase space, it satisfies the set of equations
\begin{equation}
\label{QGDeq}
\hat T_{\mu}Z(q, q_0)=0.
\end{equation}
It was also proved that the amplitude $Z(q, q_0)$ is independent on the choice of gauge conditions (\ref{u-gauge}). In the both proofs the asymptotic boundary conditions are of importance, and in the absence of the boundary conditions the proofs are not valid.

Another example of the derivation of the Wheeler -- DeWitt equation can be found in the paper by Halliwell \cite{Hall}. His starting point was the path integral with the BFV effective action. Halliwell relied upon the Fradkin -- Vilkovisky theorem which states that if the effective action is BRST-invariant then the path integral does not depend on the choice of a gauge-fixing function \cite{BFV1}. In its turn, to prove BRST invariance of the action the asymptotic boundary conditions are necessary. The validity of the boundary conditions was without doubt, and the question of the asymptotic states in quantum cosmology was not discussed. Halliwell mentioned that the Fradkin -- Vilkovisky theorem is a formal result, without referring to a particular definition of the path integral, and it would be interesting to see if the result is still true using a skeletonized definition of the path integral.

Nevertheless, the theorem is believed to hold valid, so a convenient choice of gauge conditions can be made. The choice of Halliwell was
$\dot N=0$. With this gauge choice the ghosts decouple from the other variables and the integration over the ghosts, the lapse function and their conjugate momenta can be performed in a rather simple way, again using the asymptotic boundary conditions. Then, the author argued that the resulting path integral was a solution to the Wheeler -- DeWitt equation.

Note that the gauge choice is equivalent to $N=1$. In fact, it was the same choice that had been proposed by DeWitt \cite{DeWitt}, Hawking and Page \cite{HP}, Barvinsky \cite{Barv}. Therefore, one may consider the Wheeler -- DeWitt equation in the form in which we know it as the equation under the condition $N=1$.

It was argued in \cite{SSV1,SSV2,SSV3,SSV4} that if one rejected the assumption about asymptotic states in quantum gravity one would come to a temporal Schr\"odinger equation with a gauge-dependent Hamiltonian operator, while the Wheeler -- DeWitt equation corresponds to the choice of the Arnowitt -- Deser -- Misner parametrization, the gauge conditions $N=1$ and $N_i=0$ and the additional restriction on the Hamiltonian spectrum $E=0$. This circumstance has strengthened even more the doubts if the Wheeler -- DeWitt quantum geometrodynamics is actually a gauge invariant theory.

\section{Physical states and the equivalence of different approaches to quantization}
There exist a number of approaches to quantization of gauge theories and their modifications. As we saw in the previous section, it is not a trivial task to prove their equivalence. Also, we touched upon the question how to define physical states in quantum gravity. In ordinary quantum field theory the physical states are determined by path integrals with asymptotic boundary conditions. In the Dirac approach the physical states are those annihilated by quantum constraints, and it can be argued that the both definitions are equivalent in the sense that these states do not depend on gauge and ghost variables.

The BFV approach suggest the idea of canonical quantization in extended phase space when not only physical, but also gauge and ghost variables become operators satisfying quantum commutativity relations. Then, one can define the physical states as BRST-invariant states, i.e. those annihilated by the BRST generator $\hat\Omega$ \cite{Hennaux}:
\begin{equation}
\label{BRST-gen}
\hat\Omega|\Psi^{BFV}\rangle=0.
\end{equation}
Choosing the coordinate representation for the BFV ghost variables $c^{\alpha}$, the states $|\Psi_{BFV}\rangle$ can be written as a series in the ghosts,
\begin{equation}
\label{Psi-BFV}
|\Psi^{BFV}\rangle=|\Psi^D\rangle+|\Psi^D_{\alpha}\rangle c^{\alpha}+|\Psi^D_{\alpha\beta}\rangle c^{\alpha}c^{\beta}+\cdots
\end{equation}
where $|\Psi^D\rangle$, $|\Psi^D_{\alpha}\rangle$, $|\Psi^D_{\alpha\beta}\rangle$ ``live'' in the Hilbert space of the Dirac approach not involving the ghosts. Further, it can be shown \cite{Hennaux} that a representative could be chosen among the states satisfying (\ref{BRST-gen}) which does not involve the ghosts and is annihilated by the Dirac constraints. The representative can be identified with a Dirac state vector.

The proof is based on the assumption that the classical constraints commute strongly. If they do not, from a pure theoretical point of view they can be replaced by an equivalent set of constraints having the required property. But it may be extremely difficult to find this set of constraints even in the classical theory, and it is really true for gravity. In the quantum theory we have an extra problem of ordering, because of which the same constraint may have different forms depending on additional conditions on gauge variables, as we could see by the example of the Wheeler -- DeWitt equation.

The proof of the equivalence between the Dirac quantization and canonical quantization in extended pase space includes the demonstration of equivalence of physical states, observables and scalar products in Hilbert spaces. However, the assumption about commutativity of the constraints is a keystone of the proof. The BRST generator can be constructed as a series in the ghost variables with coefficients given by generalized structure functions of constraints algebra. If the constraints commute strongly, the first and higher order structure functions would be zero, and the BRST generator would be reduced to
\begin{equation}
\label{BRST-ab}
\hat\Omega=\hat G_{\alpha}c^{\alpha},
\end{equation}
where $G_{\alpha}$ are the Dirac constraints. Because of mutual independence of the ghost variables any state satisfying (\ref{BRST-gen}) will also be annihilated by the constraints.

But at this point we meet another problem. The BRST generator is a charge that answers to global BRST symmetry, and it can be constructed according to the BFV prescription relying on the constraints algebra. On the other hand, one can apply the Noether theorem to obtain the charge making use of BRST transformation in the Lagrangian form. The both charges coincide in the case of the Yang -- Mills fields, as one could expect. But they do not coincide in the case of gravity \cite{Shest1}, and here we again encounter the peculiarities of the gravitational theory. The reason why the charge constructed by means of the Noether theorem does not coincide with the BFV charge is that the group of gauge transformations differs from the group of transformations generated by the gravitational constraints, therefore, the BRST transformations in the Lagrangian formalism are not the same as the BRST transformations for field variables in the Hamiltonian formalism. This circumstance was emphasized yet by Fradkin and Vilkovisky \cite{BFV1} who pointed out that a new type of Feynman diagrams arose. Since the new type of diagrams corresponds to a four-ghost interaction, i.e. relates to the non-physical sector of the theory, nobody paid much attention to this fact.

The BRST charge $\Omega_{NT}$ obtained in according with the Noether theorem does not have the form (\ref{BRST-ab}), so the states that satisfy the condition $\hat\Omega_{NT}|\Psi\rangle=0$ are not the Dirac states in the sense that they are not solutions to the equations
$\hat G_{\alpha}|\Psi\rangle=0$, one of which is the Wheeler -- DeWitt equation. But one cannot say that making use of the Noether theorem to construct the BRST charge is wrong. It witnesses that we deal with two nonequivalent theories with different groups of transformations, and, actually, the Wheeler -- DeWitt quantum geometrodynamics is a result of quantization of the Hamiltonian formulation for gravity proposed by Dirac and being nonequivalent to the original Lagrangian formulation of general relativity by Einstein.

So, canonical quantization in extended phase space cannot help to solve the problem of definition of physical states in quantum gravity.

\section{Discussion and conclusions}
Our notions what quantum theory of gravity must be are based on two ideas:
\begin{itemize}
\item Quantum theory of gravity must be gauge invariant;
\item Its gauge invariance is ensured by imposing the constraints on state vectors.
\end{itemize}
As we can see, the idea of gauge invariance was a lodestar for attempts to construct quantum gravity in the XXth century. Moreover, though no experimentally verified gauge theory was directly grounded on the Dirac canonical quantization, it is believed that all other approaches to quantization of gauge fields ought to be equivalent to the Dirac method. On the other side, there is no a rigorous mathematical evidence that the quantum gravitational constraints as conditions on state vectors indeed ensure gauge invariance of the theory. As we argued, the procedure of derivation of the Wheeler -- DeWitt equation, as well as the form of this equation itself, depend on additional conditions on the lapse function that restricts a choice of the reference frame.

In the present paper we have revisited some problems which quantum geometrodynamics encountered. They are closely related with the requirement of gauge invariance of this theory. Meanwhile, we have not touched upon other well-known problems, as, for example, the problem of time \cite{SS1,SS2}.

The requirement of gauge invariance in quantum geometrodynamics is stronger than in general relativity. In the latter, as Landau and Lifshitz emphasized \cite{LL}, gauge invariance implies that the laws of nature must be written in a covariant form which is appropriate to any four-dimensional system of coordinates; at the same time, the specific appearances of physical phenomena can be different in various reference frames. In a generally accepted approach, gauge invariance in quantum gravity means that the state vector (and, therefore, all physical phenomena it could determine) in no way depends on a chosen reference frame. In quantum theory the observer (by which we mean a macroscopic environment or a measuring device) plays an active role, as Wheeler repeated.

The active role of the observer can be taken into account in some mathematical realization of the Everett ``relative states'' concept \cite{Everett}. To out knowledge, the first candidate to such a realization was proposed by Barvinsky and Ponomariov \cite{BP1}. However, their proposal was ambiguous. On the one hand, they introduced a solution to the Wheeler -- DeWitt equation (\ref{BWF}) which was determined through the Faddeev -- Popov path integral with the asymptotic boundary conditions and was proved to be gauge invariant. On the other hand, the probability density to find the Universe at a given point of the Wheeler superspace is determined by the modulus squared of
\begin{equation}
\label{pr-d}
J_{\chi}\left(q, -i\frac{\partial}{\partial q}\right)\Psi(q),
\end{equation}
where $J_{\chi}$ can be called the Faddeev -- Popov operator, it explicitly depends on chosen gauge conditions $\chi^{\mu}$. The restriction of (\ref{pr-d}) to the supersurface $\Sigma$ in the superspace fixed by the gauge conditions is interpreted as a physical amplitude, or a relative state for the observer who is defined by the gauge conditions $\chi^{\mu}$. Then, the question can be posed, what quantity has s physical meaning for us as the observers inside the Universe? If it is the gauge invariant wave function $\Psi(q)$, it is not clear what could be determined by ``the physical amplitude''. But, if it is ``the physical amplitude'', the attempt of realization of the Everett concept conflicts with the idea that the physical content of the theory must be gauge invariant.

Another realization of the Everett concept was proposed in \cite{SSV1,SSV2,SSV3,SSV4}. In this approach the notion of extended phase space is exploited, and bearing in mind the features of gravity discussed above the path integral is considered without imposing the asymptotic boundary conditions. Then the Wheeler -- DeWitt equation loses its sense, and it is believed that the wave function of the Universe is a solution to the Schr\"odinger equation \cite{Shest2}. Its general solution for a model with a finite number degrees of freedom can be written as
\begin{equation}
\label{GS1}
\Psi(N, q, \theta, \bar\theta; t)
  =\int\Psi_k(q, t)\,\delta(N-f(q)-k)\,(\bar\theta+i\theta)\,dk.
\end{equation}
Here, as above, $q$ stands for physical variables, $N$ is the lapse function, and $\theta$, $\bar\theta$ are ghosts. A physical state of the Universe is determined by the function $\Psi_k(q, t)$ and this state is relative for the state of the observer in the reference frame fixed by the gauge condition $N=f(q)+k$. The function $\Psi_k(q, t)$ satisfies the Schr\"odinger equation the form of which is determined by the gauge condition. In accordance with the spirit of general relativity observers in differen reference frame can see various physical phenomena, and the price for this realization of the Everett concept is the rejection of gauge invariance of the theory.

To summarize, the idea of gauge invariance of quantum gravity, which seems to be obvious, has been confronted with many difficulties; some of these difficulties were discussed in the present paper. The path integral formalism, where gauge invariance is guaranteed by the asymptotic boundary conditions, has played the most important role in the understanding of distinctions between gravity and other gauge theories. Though it would be imprudent to state that there exists no way to construct quantum gravity in a gauge invariant manner, it is very doubtful that it could be constructed as a gauge invariant theory.
\begin{quote}
``Although there has been a lot of work in the last fifteen years\ldots\  I think it would be fair to say that we do not yet have a fully satisfactory and consistent quantum theory of gravity,''
\end{quote}
wrote Hawking in 1975 \cite{H1975}. These words are still true today. Whoever elaborate this satisfactory and consistent theory, they will have to take into account the peculiarities of the gravitational theory.


\small
\begin{thebibliography}{99}

\bibitem{DeWitt}
B. S. DeWitt,
``Quantum Theory of Gravity. I. The Canonical Theory'', {\it Phys. Rev.} {\bf 160} (1967) 1113.
\bibitem{Higgs}
P. W. Higgs,
``Integration of secondary constraints in quantized general relativity'', {\it Phys.Rev. Lett.} {\bf 1} (1958) 373.
\bibitem{Kiefer1}
C. Kiefer,
``Quantum Gravity'' (Oxford University Press, Oxford, 2007).
\bibitem{Hartle}
J. B. Hartle,
``Quantum Cosmology: Problems for the 21st Century'', in: {\it Proceedings of the 11th Nishinomiya-Yukawa Symposium},
eds. by K. Kikkawa, H. Kunitomo and H. Ohtsubo (World Scientific, Singapore, 1998).
\bibitem{Kiefer2}
C. Kiefer,
``Conceptual issues in quantum cosmology'', in: {\it Towards Quantum Gravity. Proceeding of the XXXV International Winter School on Theoretical Physics Held in Polanica, Poland, 2--11 February 1999}, ed. by J. Kowalski-Glikman (Lecture Notes in Physics, vol. 541, Springer, Berlin, Heidelberg, New York, 2000).
\bibitem{Rovelli1}
C. Rovelli,
``Quantum Gravity'' (Cambridge University Press, Cambridge, 2007).
\bibitem{Rovelli2}
C. Rovelli,
``The strange equation of quantum gravity'', {\it Class. Quantum Grav.} {\bf 32} (2015) 124005.
\bibitem{Dirac1}
P. A. M. Dirac,
``Generalized Hamiltonian dynamics'', {\it Can. J. Math.} {\bf 2} (1950) 129.
\bibitem{Dirac2}
P. A. M. Dirac,
``Generalized Hamiltonian dynamics'', {\it Proc. Roy. Soc.} {\bf A246} (1958) 326.
\bibitem{HP}
S. W. Hawking and D. N. Page,
``Operator ordering and the flatness of the Universe'', {\it Nucl. Phys.} {\bf B264} (1986) 185.
\bibitem{Barv}
A. O. Barvinsky,
``Operator ordering in theories subject to constraints of the gravitational type'', {\it Class. Quantum Grav.} {\bf 10} (1993) 1985.
\bibitem{HH}
J. B. Hartle and S. W. Hawking,
``Wave function of the Universe'', {\it Phys. Rev.} {\bf D28} (1983) 2960.
\bibitem{Hawk}
S. W. Hawking,
``The path-integral approach to quantum gravity'', in: {\it General relativity. An Einstein centenary survey},
eds. by S. W. Hawking and W. Israel (Cambridge University Press, Cambridge, 1979).
\bibitem{FP}
L. D. Faddeev and V. N. Popov,
``Feynman diagrams for Yang -- Mills field'', {\it Phys. Lett.} {\bf B25} (1967) 29.
\bibitem{Fadd}
L. D. Faddeev,
``The Feynman integral for singular Lagrangians'', {\it Theor. Math. Phys.} {\bf 1} (1969) 1.
\bibitem{BFV1}
E. S. Fradkin, G. A. Vilkovisky,
``Quantization of relativistic systems with constraints'', {\it Phys. Lett.} {\bf B55} (1975) 224.
\bibitem{BFV2}
I. A. Batalin, G. A. Vilkovisky,
``Relativistic S-matrix of dynamical systems with boson and fermion constraints'', {\it Phys. Lett.} {\bf B69} (1977) 309.
\bibitem{BFV3}
E. S. Fradkin, T. E. Fradkina,
``Quantization of relativistic systems with boson and fermion first- and second-class constraints'', {\it Phys. Lett.} {\bf B72} (1978) 343.
\bibitem{BV}
I. A. Batalin, G. A. Vilkovisky,
``Gauge algebra and quantization'', {\it Phys. Lett.} {\bf B102} (1981) 27.
\bibitem{BP1}
A. O. Barvinskiy and V. N. Ponomariov,
``Canonical quantization of gravity and quantum geometrodynamics'', {\it Izv. vuzov, Fizika} {\bf 3} (1986) 37.
\bibitem{BP2}
A. O. Barvinskiy and V. N. Ponomariov,
``Quantum geometrodynamics: the path integral and the initial value problem for the wave function of the Universe'', {\it Phys. Lett.} {\bf B167} (1986) 289.
\bibitem{Hall}
J. J. Halliwell,
``Derivation of the Wheeler -- DeWitt equation from a path integral for minisuperspace models'', {\it Phys. Rev.} {\bf D38} (1988) 2468.
\bibitem{SSV1}
V. A. Savchenko, T. P. Shestakova and G. M. Vereshkov,
``Quantum Geometrodynamics of the Bianchi IX model in extended phase space'', {\it Int. J. Mod. Phys.} {\bf A14} (1999) 4473.
\bibitem{SSV2}
V. A. Savchenko, T. P. Shestakova and G. M. Vereshkov,
``The exact cosmological solution to the dynamical equations for the Bianchi IX model'', {\it Int. J. Mod. Phys.} {\bf A15} (2000) 3207.
\bibitem{SSV3}
V. A. Savchenko, T. P. Shestakova and G. M. Vereshkov,
``Quantum Geometrodynamics in extended phase space - I. Physical problems of interpretation and mathematical problems of gauge invariance'',
{\it Grav. Cosmol.} {\bf 7} (2001) 18.
\bibitem{SSV4}
V. A. Savchenko, T. P. Shestakova and G. M. Vereshkov,
``Quantum Geometrodynamics in extended phase space - II. The Bianchi IX model'', {\it Grav. Cosmol.} {\bf 7} (2001) 102.
\bibitem{Hennaux}
M. Hennaux,
``Hamiltonian form of the path integral for theories with a gauge freedom'', {\it Phys. Rep.} {\bf 126} (1985) 1.
\bibitem{Shest1}
T. P. Shestakova,
``The role of BRST charge as a generator of gauge transformations in quantization of gauge theories and gravity'',
{\it Tomsk State Pedagogical University Bulletin} {\bf 153} (2014) 224.
\bibitem{SS1}
T. P. Shestakova and C. Simeone,
``The problem of time and gauge invariance in the quantization of cosmological models. I. Canonical quantization methods'',
{\it Grav. Cosmol.} {\bf 10} (2004) 161.
\bibitem{SS2}
T. P. Shestakova and C. Simeone,
``The problem of time and gauge invariance in the quantization of cosmological models. II. Recent developments in the path integral approach'', {\it Grav. Cosmol.} {\bf 10} (2004) 257.
\bibitem{LL}
L. D. Landau and E. M. Lifshitz,
``The Clssical Theory of Fields'' (Oxford, 1975).
\bibitem{Everett}
H. Everett,
`` 'Relative state' formulation of quantum mechanics'', {\it Rev. Mod. Phys.} {\bf 29} (1957) 454.
\bibitem{Shest2}
T. P. Shestakova,
``Is the Wheeler - DeWitt equation more fundamental than the Schrödinger equation?'', {\it Int. J. Mod. Phys.} {\bf D27} (2018) 1841004.
\bibitem{H1975}
S. W. Hawking,
``Particle Creation by Black Holes'', {\it Commun. Math. Phys.} {\bf 43} (1975) 199.
\end{thebibliography}
\end{document}